\documentclass[prc,amsmath,amssymb,dvips,preprint]{revtex4} 
\usepackage{epsfig} 
\usepackage{bm} 
\usepackage{amsmath}
\usepackage{rotating}
 
\newcommand{\be}{\begin{equation}} 
\newcommand{\ee}{\end{equation}} 
\newcommand{\bea}{\begin{eqnarray}} 
\newcommand{\eea}{\end{eqnarray}}

\begin{document} 
 
\title{Energy spectrum and effective mass using a non-local 3-body interaction}
 
\author{Alexandros Gezerlis$^1$ and G.~F. Bertsch$^{1,2}$ } 
\affiliation{$^1$Department of Physics, University of Washington, Seattle, WA 98195--1560 USA} 
\affiliation{$^2$Institute for Nuclear Theory, University of Washington, Seattle, WA 98195--1560 USA} 
 
\date {\today} 
 
\begin{abstract} 
We recently proposed a 
nonlocal form for the 3-body induced interaction that is
consistent with the Fock space representation of interaction 
operators but leads to a fractional power dependence on the density. Here 
we examine the implications of the nonlocality for the excitation
spectrum.  In the two-component weakly interacting Fermi gas,
we find that it gives an
effective mass that is comparable to the one in many-body perturbation theory.
Applying the interaction to nuclear matter, it predicts  a
huge enhancement to the effective mass.  Since the
saturation of nuclear matter is partly due to the induced 3-body 
interaction, fitted functionals should treat the effective mass
as a free parameter, unless the two- and three-body contributions
are determined from basic theory.
\end{abstract} 
 
 
\maketitle 

Zero- and finite-range nuclear energy-density functionals
have a long history and a successful track record, allowing
the description of heavy nuclei without region-specific 
parametrizations.\cite{Bender:2003}
The most popular functionals use interactions that depend
on fractional powers of density, which causes serious problems when one
tries to extend the theory to include correlations
\cite{Duguet:2003,Robledo:2007,Duguet:2009}.  
Ideally,  to avoid these problems the
effective theory should be based on a Fock-space Hamiltonian
operator.  As a partial solution, one can consider energy functionals
of integral powers of the density; there have been a number of
attempts to construct functionals of this kind
\cite{Baldo:2010,Erler:2010}.

With this in mind,  we recently 
proposed a nonlocal effective three-body interaction 
that achieves a fractional dependence on density using only integral
powers of the density matrix \cite{Gezerlis:2010a}. This was derived using the many-body perturbation
theory of the dilute, weakly interacting Fermi gas. By construction, the
interaction gives the correct Lee-Yang contribution \cite{Lee:1957} to the Fermi-gas energy 
to order $\rho^{7/3}$.  The interaction was validated for finite
systems in a harmonic trap by comparing with numerically accurate calculations performed
by the Green's Function Monte Carlo method.
At very weak coupling, the new operator
led to results that are identical with the Lee-Yang dependence, while 
for stronger coupling the contribution of the new 3-body operator turned out to be
more repulsive than in Lee-Yang (though with
the same power-law behavior), thus providing
a more accurate description of the microscopic simulation. 

Using the new interaction, the internal energy of the dilute
Fermi gas can be expressed in terms of the one-body density matrix as:
\begin{equation}
\label{E_first} 
\begin{split}
E &= \frac{\hbar^2 k_F^2}{m}\int d^3r_1\,
\left(\frac{\nabla_{r_1}\cdot\nabla_{r_2}}{2}\rho({\bf r_1, r_2})|_{
{\bf r_1 = r_2}} + 
4\pi a \rho_\downarrow({\bf r_1},{\bf r_1})\rho_\uparrow({\bf r_1},{\bf r_1})
\right)\\
 &+ C\int d^3 r_1 d^3r_2\,\frac{\rho_\uparrow({\bf r_1, r_2})
\rho_\downarrow({\bf r_1, r_2})\rho({\bf r_1, r_2})}{|{\mathbf r_1}-{\mathbf
r_2}|}.
\end{split}
\end{equation}
The subscript $i$ on $\rho_i$ denotes the spin state, 
$\rho$ without a subscript is the total density.  Also, if $a$ is the scattering length 
associated with the two-body interaction,
$C$ is a constant proportional to $a^2$.  The value of $C$ was derived
in Ref. \cite{Gezerlis:2010a} by demanding that the
formula reproduce the Lee-Yang energy in the uniform Fermi gas.  The
energy $E$ or energy density $\cal E$ is given by 
\be 
\label{LY} 
\frac{E}{A} = \frac{\cal E}{\cal \rho} = \frac{\hbar^2 k_F^2}{2 m}\left(
\frac{3}{5} + \frac{2}{3\pi}ak_F + \frac{4}{35\pi^2} \left ( 11 - 2 \ln2
\right ) \left ( ak_F \right )^2\right)~.  
\ee 
It is convenient for later use to rederive from Eq. (1) the formula for $C$,
which was originally derived from the perturbation theory in a
momentum space representation.  We insert in Eq. (1) the free Fermi
gas density matrix and drop one of the integrals to get the energy 
density.  The Fermi gas density matrix 
only depends on the relative coordinate
${\bf r} = {\bf r_1}- {\bf r_2}$ and can be written
\begin{equation}
\label{rho_0} 
\rho_{0i}({\bf r}) = \int_0^{k_{F}} \frac{d^3k}{(2 \pi)^3}
e^{i {\bf k} \cdot {\bf r}} = \rho_{0i}(0) F(k_{F}r)
\end{equation}
where $F(x) = 3 j_1(x)/x$.  
The integral to be evaluated may be expressed
\begin{equation}
\label{ed_first} 
\frac{E_3}{A} = C\frac{24\pi^3}{ k_F^5} \rho_{0i}^3(0) \int_0^\infty
x  dx F^3(x)
\end{equation}
The integration can be performed analytically; the final result for the
strength parameter $C$ is
\be
\label{C}
C = {\hbar^2 a^2\over m}{64\pi (11 - \ln2)\over 3(92-27\ln3)} 
\ee

We now calculate the single-particle energy with functional Eq. (1) and the
value of $C$ fixed by Eq. (\ref{C}).  The density matrix with
a particle added to the Fermi sea is
\be
\label{rhok}
\rho_i({\bf r)} = 
\rho_{0i}({\bf r}) + \rho_{ki} e^{i {\bf k} \cdot {\bf r}} 
\ee
The second term
represents a particle of momentum $k$ in spin state $i$; the coefficient 
$\rho_{ki}$ has dimensions of density.  With this definition the single-particle
energy may be computed as
\be
\varepsilon_i(k) = \frac{d {\cal E}}{d \rho_{ki}} \Big|_{\rho_{ki} = 0}.
\ee 
Carrying out the differentiation on the energy expression Eq. (\ref{E_first}), the
first term gives the usual kinetic energy and the second term is
independent of $k$.  The third term is rather complicated. Assuming
equal populations of spin up and spin down in $\rho_0$, 
the derivative  
is given by the integral:
\be
\label{sp_first}
\varepsilon_3(k) = 3C  \int d^3 r\, e^{i {\bf k} \cdot ({\bf r})} 
\frac{\rho_{0i}^2({\bf r})}{r}= 3 C {4 \pi\over k_F^2} \rho_{0i}^2(0)
\int_0^\infty dx ~ x\,\, j_0 \Big( {k\over k_F} x \Big) F^2(x) .
\ee
The factor of 3 is a direct consequence of the spin structure of the 
numerator in the third term of Eq. (\ref{E_first}). The integral 
can also be expressed analytically:
\begin{equation}
\begin{split}
\int_0^\infty dx ~ x\, j_0 ( y x) F^2(x) &= \frac{3}{160}\Big[ 4 (22+y^2) + \frac{(y-2)^3}{y}(y^2 + 6y + 4)\log(2-y)\\
 &- 2y^2(y^2-20)\log y + \frac{(y+2)^3}{y}(y^2 - 6y + 4)\log(2+y) \Big] ,
\end{split}
\end{equation}
where $y=k/k_F$.
The 3-body contribution to the single-particle energy $\varepsilon_3$ is plotted in
Fig. 1, with the dimensionful factors divided out.  
\begin{figure}[t]
\vspace{0.5cm}
\begin{center}
\includegraphics*[width=0.46\textwidth]{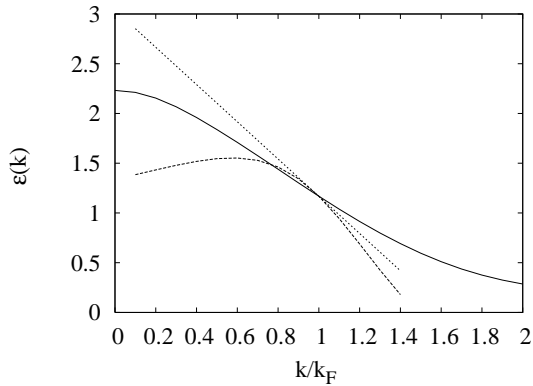}
\caption{Single-particle energy in the dilute Fermi gas, normalized to
$\hbar^2 a^2 \rho_{0i}^2(0)/m k_F^2$.  The solid line shows result using 
the effective 3-body interaction, Eq. (\ref{sp_first}).  The dashed line
shows the contribution to the quasiparticle energy obtained by 
Galitskii.  The dotted line is the slope of the Galitskii expression.
}
\label{fig:modi}
\end{center}
\end{figure}
Galitskii's expression for the real 
part of the quasiparticle energy \cite[Eq. (34)]{Galitskii:1958} is plotted
with the same normalization on the graph.
The perfect agreement of the two at the Fermi momentum is not accidental:
the single-particle energy
at the Fermi surface is identical to the chemical potential $\mu$, which 
can be extracted from the interaction energy by the formula
$\mu = \partial {\cal E}/ \partial \rho$.  Since we fit the
total 3-body interaction energy to the dilute Fermi gas, the
chemical potential must agree as well.

The momentum-dependence
of the single-particle energy gives rise to an effective mass
$m^*$ for the quasiparticle spectrum,
\be
\label{eq:effmass}
\frac{m^*}{m} = \left ( 1 + {m\over \hbar^2 k_F}{\partial \epsilon_3(k)\over \partial k}\Big|_{k=k_F} \right )^{-1}
\ee
The derivative in this expression is negative, implying that the effective
mass will be larger than $m$.  Fig. 1 also shows the derivative
for Galitskii's quasiparticle energy, as the straight 
line (see also Ref. \cite{Fetter:1971}). We note
that the slope for the 3-body single-particle energy is smaller, implying
less of an effective-mass enhancement. Even so, the two results
are close enough in magnitude to motivate the application of the 
new operator to a nuclear energy functional.

As stated in the introduction, our main interest is to find
an improved effective Hamiltonian for nuclear structure theory.
There is no reliable low-density expansion in the nuclear many-body problem,
and in fact one must impose some length scale in the interactions to
avoid collapse.  Nevertheless, in some formulations there will be
a contribution to saturation coming from the Pauli effects that we
are concerned with here.  To assess the importance of the nonlocality,
we take $C$ as an adjustable parameter to be fitted in the functional,
similar to the parameter $t_3$ of the Skyrme interaction.
The counting
of the contributing graphs is different in the four-component Fermi
system than in the two-component case treated by Galitskii, but the
scaling between the total energy and the single-particle energy remains
the same under plausible assumptions about the spin-isospin character
of the interaction.
Thus we may use the same formulas, only remembering
that in the nuclear context $\rho_{0i}$ is the density associated
with a specific spin-isospin projection, e.g. neutrons with spin up.

\begin{table}[t]
\caption{Contributions to the energy of $^{208}$Pb in density functional
theory. The numbers for
the Skyrme Ska and Gogny D1S functionals were obtained with the ev8 code
\cite{Bonche:2005} and the HFBaxial code \cite{robledo},
respectively.}
\begin{center}

\begin{tabular}{|l| rr|}
\hline
 & Ska & D1S \\
\hline
Kinetic & 3863 & 3920\\
Coulomb direct/exchange & 831/-31 & 832/-31\\
Spin-orbit & -97 & -105\\ 
Central 2B  & -12480 & -12783 \\ 
$t_3$ &  6274 & 6530\\
Total & -1640 & -1637\\
\hline
\end{tabular}

\end{center}
\label{table1}
\end{table}

While we cannot calculate $C$, we can at least put a bound on its value
using the magnitude of the 3-body interaction energy that is obtained
from phenomenological energy functionals.  With our form for the
interaction, the relation between the 3-body energy  and the effective
mass is
\be
\label{mstar}
\frac{m^*}{m} = \left ( 1 +d \frac{E_3/A}{\hbar^2 k_F^2 / (2m)} \right )^{-1}
\ee
where $d\approx -1.32$, and the two-body contribution has been omitted.

To see what the scale of the effect would be, we show in Table 1 
the various contributions to the energy of $^{208}$Pb found using the
Ska Skyrme functional and the D1S Gogny functional.  Both these functionals
have the same $\rho^{1/3}$ density-dependent interaction as in the
Lee-Yang expansion.  One sees that the decomposition into 
the two-body and three-body contributions is quite similar, although the
two-body interactions have a very different construction.
Eq.
(\ref{mstar}) gives a 
negative effective mass for both functionals, which is of course unphysical.
The two-body
nonlocality gives a contribution of the opposite sign, but not enough
to produce an effective mass in the physical range ($m^*/m\sim 1$).
As mentioned earlier, there
must be other 3-body contributions containing intrinsic length
scales in order to achieve nuclear saturation. However, unless the nonlocalities
can be calculated in detail, it does not seem feasible to derive
a theoretical effective mass to be used with an effective Hamiltonian.
The extreme sensitivity to the induced 3-body interaction suggests
that the effective mass may need to be an unconstrained free
parameter when constructing an effective Hamiltonian for mean-field
theory and its extensions.
   
In summary, we have applied our newly proposed non-local effective 3-body
operator to the study of the single-particle excitation spectrum, both
at weak coupling and at strong coupling. At weak coupling we see that
the new operator has similar behavior to that found by Galitskii. 
We also applied the new operator to the nuclear case. 
The effects pointed to are very large, implying that
the effective mass cannot be simply taken to be reduced from
the bare mass based on mean-field theory: as long as no
dependable {\it ab initio} results are available, the effective
mass should also be treated like an undetermined parameter.

We would like to thank L.~Robledo for providing us with the energies of
$^{208}$Pb for the Gogny D1S interaction.
This work was supported by DOE Grant Nos. DE-FG02-97ER41014 and DE-FG02-00ER41132.

 
\end{document}